\documentclass[twocolumn]{revtex4}
\usepackage{graphics,epsfig,graphicx}
\usepackage{color}
\usepackage{makeidx}
\usepackage{latexsym}

\begin{document}


\title{A single polymer chain as an organic quantum wire: optical evidence of a purely 1D density of states}

\author{ F. Dubin$^*$, J. Berr\'{e}har, R. Grousson, T. Guillet,  C. Lapersonne-Meyer,\\ M. Schott, V. Voliotis \\
{\em Groupe de Physique des Solides, UMR 7588 of CNRS,} \\
{\em Universit\'{e}s Pierre et Marie Curie et Denis Diderot, Tour 23, 2 place Jussieu,} \\
{\em 75251 Paris cedex 05, France }}


\begin{abstract}
The excitonic luminescence of an isolated polydiacetylene polymer
chain in its monomer matrix is studied by micro-photoluminescence.
These chains behave as perfect 1D semiconducting wires with the
expected 1/$\sqrt{E}$ density of states between 5 and 50 K. The
temperature dependence of the homogeneous width is explained by
interaction with longitudinal acoustic phonons of the crystal in
the range of temperature explored. The optical phonons of the
chain which are involved in the vibronic transitions are found to
have coherence times ranging from 300 to 600 fs.
\end{abstract}

\maketitle

Excitons are the dominant optically accessible states in
conjugated polymers. They are usually discussed in terms of single
discrete states, as in an isolated molecule. The exciton band
structure and density of states (DOS) are usually not considered,
unlike in inorganic semiconductors for which the exciton energy
dispersion and DOS are important properties. For the past decade,
an important effort has been made in order to obtain a 1D
semiconducting system exhibiting a $1/\sqrt{E}$ DOS. Such a DOS
has neither been observed in the quantum wires which have been
realized \cite{Kapon}, nor in organic "1D" J aggregates
\cite{Potma}. In the present letter, we report
micro-photoluminescence ($\mu$-PL) experiments on a single
isolated poly-3BCMU \cite{formule3B} red chain and show that to
properly describe excitons in this system the actual band
structure must be considered. We show as well that the DOS is the
one of perfect 1D systems with the expected 1/$\sqrt{E}$ singular
variation. The temperature dependence of the homogeneous width is
explained by interaction with acoustic phonons \cite{Dubin2002},
and the values found for the chain optical phonon coherence times
are also presented.\

PDA chains diluted in 3BCMU crystal monomer matrix exist in two
electronic structures so called "red" and "blue" phases. Both
types of chains exhibit an excitonic resonance fluorescence
\cite{Lecuiller98}. Blue chain fluorescence is very weak with a
quantum yield of $10^{-4}$ whereas red chains have a high
fluorescence quantum yield of 0.3 at 15 K \cite{Lecuiller99}. The
emitting species of such chains are excitons with a large binding
energy of 0.5 eV \cite{Horvath}. The luminescence spectrum of red
chains exhibits an intense zero phonon line and several much
weaker vibronic replicas. The zero phonon line is centered at 2.28
eV at low temperature. The two main vibronic peaks correspond to
the stretching of the C$=$C and C$\equiv$C bounds and will now be
denoted by D and T respectively. These two lines are centered at
2.09 and 2.01 eV and are due to radiative recombination with
emission of a chain optical phonon with the appropriate momentum
\cite{Lecuiller98}. The very high dilution of red chains and their
high fluorescence yield allow the study of a single red chain by
$\mu$-PL experiments. The zero phonon line was studied giving
access to the homogeneous linewidth $\Gamma_{0}$ and its
temperature dependence. It also brought evidence on long range
energy transfer along the chain \cite{Guillet2001}. In the present
paper this study is extended to the vibronic emission lines.

\begin{figure}
\begin{center}
\mbox{\includegraphics[width=8 cm,height=6 cm]{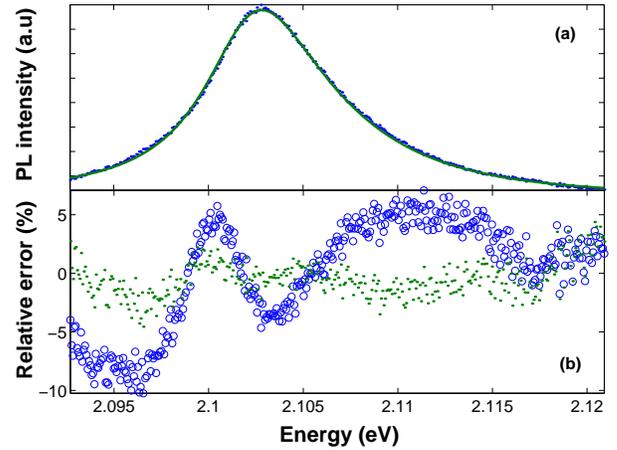}}
\caption{(a) Data and  fit using a 1D DOS of the D peak centered
at 2.102 eV for T=43 K. (b) Comparison of the error (relative to
the maximum signal value) between the fit function and the data
for the 1D model (points) and for the 2D model (open circles).}
\label{raievib}

\end{center}

\end{figure}

Macro-PL experiments on an ensemble of red chains
\cite{Lecuiller2002} show that up to 50 K the exciton lifetime
increases with temperature while the fluorescence rate decreases.
This suggests that non-radiant states are populated by thermal
excitation of radiant ones, a known process in exciton bands of
semiconductor quantum wires. Nevertheless, in organic materials
the exciton is usually considered as a single excited state. The
analysis of the vibronic peaks line-shapes remove this ambiguity
since they reflect the DOS in a band description.\

The 3BCMU single crystals analyzed here were identical to the ones
described in \cite{Guillet2001}. The average concentration of red
chains is smaller than $10^{-8}$ in weight, allowing the study of
a single chain. The $\mu$-PL experimental setup is the same as in
\cite{Guillet2001}. The excitation wavelength of the $Ar^{+}$
laser was chosen at 497 nm, nearly resonant to the D (Double bound
stretch) absorption line. The excitation power was of the order of
one $\mu$W to keep the measurement in the low excitation regime,
i.e. with at most one exciton per chain.\

The D emission line at T=43 K is shown in Figure \ref{raievib}a.
The line-shape is clearly non lorentzian, this emission can then
not be that of a single state. In Fig.\ref{raievib}a, a fit barely
distinguishable from the experimental data is presented as well,
and the corresponding error is given in Fig.\ref{raievib}b. This
fit is based on 3 assumptions:
\begin{itemize}
\item The excitons in the chain are at
thermodynamic equilibrium with the surrounding medium. Their
energy distribution then follows a Maxwell-Boltzmann distribution
characterized by the crystal's overall temperature.
\item The energy dispersion of the optical phonons generated in
the vibronic emissions is very small compared to the one of the
excitons and is neglected. Since the effective mass of the exciton
is found to be $\approx$ 0.3 $m_{0}$ \cite{Lecuiller2002} ($m_{0}$
is the bare mass of an electron), this hypothesis seems very
reasonable.
\item The transition matrix elements between all the initial \textbf{k} exciton states and  the
final state (emission of one photon and one chain optical phonon)
are equal (including the \textbf{k}=0 state). The contribution of
the initial \textbf{k} state to the overall homogeneous width  is
the same for all \textbf{k}.
\end{itemize}

The fitting function $f(E)$ is the convolution of the homogeneous
lorentzian profile and the population of emitting states
(\ref{fitfunc}). This population is the product of the DOS by the
occupation probability. In (\ref{fitfunc}) $E_{0}$ is the
lorentzian's center position and $\Gamma_{vib}$ its half width.
$\it{A}$ is a constant including the amplitude of the lorentzian
and the constant parameters of the DOS. The only relevant
parameter in the fitting routine is $\Gamma_{vib}$ (e.g.
$\Gamma_{D}$ for the D line) which dependence with temperature
will be discussed.

\begin{equation}\label{fitfunc}
  f(E)=\frac{A}{(E-E_{0})^{2}+\Gamma_{vib}^{2}}\star
  \left(\exp(-\frac{E-E_{0}}{k_{B}T}).DOS \right)
\end{equation}

A quantitative 1D fit, i.e with a $(E-E_{0})^{-1/2}$ DOS, is
obtained for all vibronic lines at all temperature studied (5-50
K). Fitting with a 2D DOS is always worse. The 2D DOS corresponds
to the lowest dimensionality non singular density of states, and
has been found to fit data on J aggregates \cite{Potma}. The
fitting error for the 1D and 2D model are presented in
Fig.\ref{raievib}b, and show the very good accuracy of our model
when a 1D DOS is used.

\begin{figure}
\begin{center}
\mbox{\includegraphics[width=8 cm,height=7 cm]{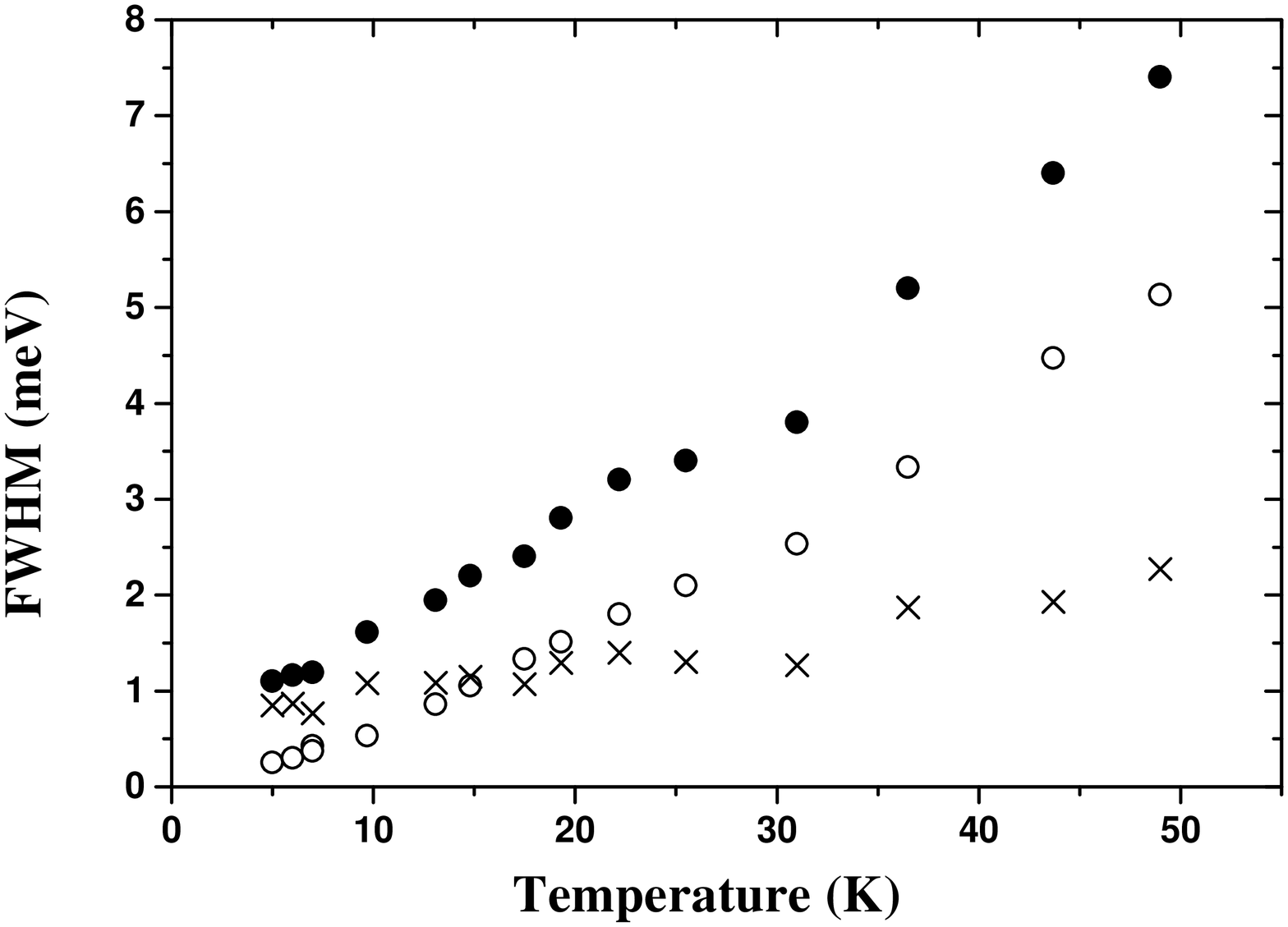}}
\caption{$\Gamma_{D}$ (filled circles) and $\Gamma_{0}$ (open
circles) versus temperature. The crosses represent the difference
between $\Gamma_{D}$ and $\Gamma_{0}$.} \label{W(T)}
\end{center}
\end{figure}

\begin{figure}
\begin{center}
\mbox{\includegraphics[width=7 cm,height=4 cm]{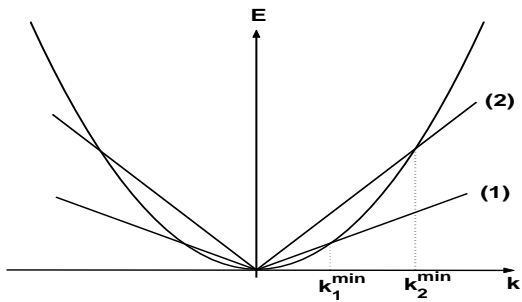}}
\caption{Dispersion relations of the excitons and LA phonons. (1)
and (2) are the dispersion curves of LA phonons of the crystal and
LA phonons confined on the chain respectively. In the first case
the sound velocity is assumed isotropic and is $2.5\ 10^{3}$
m.$s^{-1}$. In the latter case the sound velocity is\\ $5.5\
10^{3}$ m.$s^{-1}$. $k_{1,2}^{min}$ denotes the scattering
thresholds in each case.} \label{scatt}
\end{center}
\end{figure}

\begin{figure}
\begin{center}
\mbox{\includegraphics[width=6 cm,height=6 cm]{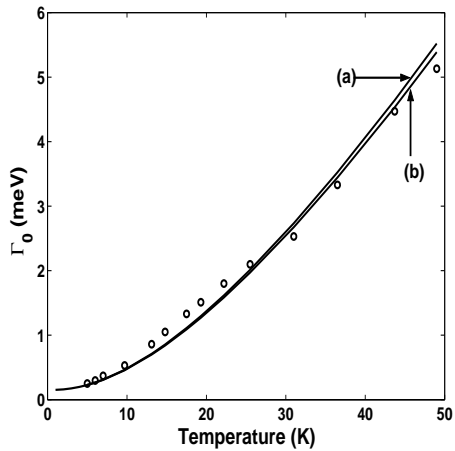}}
\caption{Experimental data (open circles) and model of the
broadening of $\Gamma_{0}$ with temperature for an exciton Bohr
radius of 10 and 20 \AA ((a) and (b) respectively) and an
effective mass of 0.3 $m_{0}$. In both cases, the sum of the
deformation potential for the valence and conduction band is 6
eV.} \label{fitgamma0}
\end{center}
\end{figure}

$\Gamma_{D}$(T) and $\Gamma_{0}$(T) are presented in
Fig.\ref{W(T)}. The present data on $\Gamma_{0}$(T) confirm
previous measurements presented in \cite{Guillet2001}:
$\Gamma_{0}$ is 250 $\mu$eV at 5 K, then increases  linearly up to
35 K. The model used to account for the variation with temperature
of $\Gamma_{0}$ assumes that the dominant process is the
absorption of a single acoustic phonon (LA phonon) of the crystal
\cite{Dubin2002}. It is an adaptation of the deformation potential
theory to a 1D system such as poly-3BCMU (see for instance  Oh and
Singh \cite{Singh2000}). Due to the energy and momentum
conservation rules, LA phonons confined on the polymer chain can
only make exciton transitions between \textbf{k}=0 and
\textbf{k}=$\textbf{k}_{2}^{min}$ (see Fig. \ref{scatt}). This
does not agree with the experimental observation that all
\textbf{k} states emit within an energy range $k_{B}$T. On the
contrary, exciton scattering by crystal LA phonons connect
\textbf{k}=0 and a continuum of \textbf{k} states with \textbf{k}
$> \textbf{k}_{1}^{min}$ ($\textbf{k}_{1}^{min}$ is defined in
Fig.\ref{scatt}) since only the component parallel to the chain of
the total momentum must be conserved. Further scattering can
populate the rest of the band. Therefore, we have considered
interactions with LA phonons of the crystal. This process
quantitatively explains the variation of $\Gamma_{0}$(T) in the
range of temperature explored (see Fig. \ref{fitgamma0}). The
sound velocity is assumed isotropic in the crystal and is taken as
2.5\ $10^{3}\ m.s^{-1}$ as in another diacetylene \cite{vs}. This
is a typical value for molecular crystal.\

Besides the chosen sound velocity, the calculation involves three
parameters: the exciton Bohr radius, its effective mass, and
$(D_{c}+D_{v})$ the sum of the deformation potential for the
conduction and valence band respectively. Experiments and theory
indicate that the Bohr radius of the exciton on PDA chains is
between 10 and 20 \AA\ \cite{Sushai,Horvath}. The chosen value for
this parameter is not critical as shown in Figure \ref{fitgamma0}.
From the temperature dependence of radiative lifetime, an exciton
effective mass ($m^{*}_{X}$) of approximately 0.3 $m_{0}$ is
deduced with a large uncertainty \cite{Lecuiller2002}. To
$m^{*}_{X}=0.3\ \pm 0.1\ m_{0}$ correspond $(D_{c}+D_{v})$ values
equal to $6.1\ \mp 0.8$\ eV, which are typical deformation
potentials. Since the scattering rate of excitons by LA phonons
goes to zero at 0 K, a constant value corresponding to the
residual linewidth at 0 K has been added. Its fitted value is
150~$\mu$eV, much larger than the contribution of the effective
lifetime of the exciton at 1 K (approximately 6 $\mu$eV)
\cite{Lecuiller2002}. This remains an open question and will be
the subject of further analysis.

The inelastic processes thermalizing the excitons in their band
are thus either the emission or absorption of LA phonons in the
temperature range studied \cite{Dubin2002}. The absorption is the
slowest process and has a characteristic time of 2 ps or less (see
the widths in Fig.\ref{W(T)}). The excitons are then at
thermodynamic equilibrium since their effective lifetime is over
100 ps in the range of temperatures studied \cite{Lecuiller2002}.\

In Fig.\ref{W(T)}, one notes that $\Gamma_{D}$ is 3 times larger
than $\Gamma_{0}$ at low temperature. Let us assume that the width
of a vibronic line is the sum of the terms resulting from the
initial and final state. The contribution of the initial state is
$\Gamma_{0}$. The one from the final state represents the
coherence time of the optical phonon emitted in the recombination
which is slowly decreasing with temperature (see for instance the
Fig.\ref{W(T)} for the D line). One would expect that the
coherence time so obtained is not the same for all vibronic peaks.
This is indeed observed: a coherence time of approximately 300 fs
is found for the D and T lines while in another vibronic line
centered at 2.15 eV, distinctly narrower, a coherence time of 600
fs is found. These different values are in agreement with previous
measurements made in another PDA by Chen et al. using a completely
different method (CARS) \cite{Chen1,Chen2}.\

Fitting the line-shape of the vibronic peaks has enabled us to
deduce that our system exhibits a $1/\sqrt{E}$ DOS. Let us
associate this conclusion with the analysis of the exciton decay
time presented in \cite{Lecuiller2002}. Indeed, the radiative
lifetime of excitons increases as $\sqrt{T}$ for T up to 80 K at
least, as it is expected for an ideal 1D wire \cite{Citrin92}.
Therefore, localization effects do not seem to affect the exciton
lifetime and DOS, contrary to the case of inorganic semiconductor
quantum wires \cite{Kapon}. Time resolved $\mu$-PL experiments are
then planned to further study exciton transport and localization
on a single chain.\

To summarize, we have presented microfluorescence experiments
performed on a single conjugated polymer chain in its crystalline
matrix. The zero phonon emission line is lorentzian while the
vibronic ones are asymmetric. Fitting the line-shape of these
vibronic peaks shows that the chain is a one dimensional system
which has to be described by an excitonic band with a $1/\sqrt{E}$
DOS. Furthermore,  the variation of $\Gamma_{0}$ in temperature is
explained by interactions with longitudinal acoustic phonons of
the crystal in the range of temperature studied. The chain optical
phonons involved in the vibronic emissions are found to have
coherence times ranging from 300 to 600 fs at low temperature.\

The authors are thankful to P. Lavallard and G. Weiser for very
helpful discussions and to J. Kovensky for synthesizing the
diacetylene monomer. This research is supported by CNRS within the
program "nano-objet individuel", and by a grant from University
Denis Diderot.

\noindent

[*] Corresponding author: dubin@gps.jussieu.fr


\begin{thebibliography}{99}
\bibitem{formule3B} The general formula of PDA is
$( =\! CR \! - \! C \! \equiv \! C \! - \! CR' \!=)_{n}$. In
3BCMU, the side groups R and R' are identical, with the following
molecular formula: -$(CH_{2})_{3}OCONHCH_{2}COOC_{4}H_{9}$
\end{thebibliography}
\end{document}